\documentclass[%
 preprint,
 amsmath,amssymb,
 aps,
 pre,
showkeys
]{revtex4-2}

\usepackage{graphicx}% Include figure files
\usepackage{dcolumn}% Align table columns on decimal point
\usepackage{bm}% bold math
\usepackage{hyperref}% add hypertext capabilities
\usepackage{xcolor}
\usepackage{ctable}
\begin{document}

\title{Surface tension of Bose-Einstein condensate at finite temperature}% Force line breaks with \\

\author{ Nguyen Van Thu}
\affiliation{Department of Physics, Hanoi Pedagogical University 2, Hanoi, Vietnam}
\email[]{nvthu@live.com}

%\email[]{jonas.berx@kuleuven.be}

\date{\today}% It is always \today, today,

\begin{abstract}
We examine the influence of nonzero temperature on a bounded surface of a dilute Bose gas confined by a hard wall, utilizing the Gross-Pitaevskii theory and the Bogoliubov-de Gennes equation to describe surface excitations. The theoretical calculations are compared with experimental data for liquid helium II, demonstrating excellent agreement. An empirical relation is found for the temperature-dependence of the surface tension. Furthermore, our findings indicate that the contribution of surface excitations remains below 0.7\% in the most of recent experiments on Bose-Einstein condensate.
\end{abstract}


\maketitle

%\tableofcontents
\section{Introduction\label{sec1}}

The surface tension of classical as well as quantum liquids represents the free energy per unit area of their surface. It is a crucial parameter that plays a key role in determining the properties of any liquid \cite{Rowlinson2002}. In the wetting phenomenon, surface tension governs an ability of a liquid to spread upon contact with a substrate  \cite{Rowlinson2002,Gennes2009}.

In the study of wetting phenomena in Bose-Einstein condensates (BECs), the possibility of wetting induced by a BEC adsorbed on a hard wall was first investigated in 2004 \cite{Indekeu2004}. Subsequently, the wetting phase diagram of a BEC adsorbed on both hard and soft walls was also examined in Refs. \cite{VanSchaeybroeck2015,Indekeu2015,Thu2016,DuyThanh2024}. However, to the best of our knowledge, experimental verification of the wetting phenomenon in BECs has not yet been conducted and remains an open challenge. Several factors have been identified as potential obstacles, the foremost being the complexities associated with the production of a hard wall \cite{Indekeu2023}, despite significant advancements in cooling technology \cite{Navon2021}. A recently proposed alternative solution involves the use of a ternary BEC, wherein the hard wall is replaced by the interface between two condensed components, while the third component may wet this interface \cite{Indekeu2023,Jimbo2021}. Another theoretical prediction suggests that the static properties of the prewetting state should be considered \cite{DuyThanh2024}. A common characteristic of existing studies on wetting in BECs is the reliance on the Gross-Pitaevskii theory, which inherently assumes that the system is at zero temperature. This approximation neglects the role of temperature, despite the fact that absolute zero has never been achieved, even with state-of-the-art technologies. Nevertheless, temperature may play a crucial role in wetting phenomena, yet, to the best of our knowledge, its effects have never been explicitly addressed.

The main aim of this study is to investigate the influence of nonzero temperature on the surface tension of a single BEC confined by a hard wall and to compare the findings with available experimental data for helium isotopes. In particular, the study focuses on assessing the contribution of surface excitations to surface tension. This effect may have profound implications for the feasibility of the wetting phenomenon in BECs.

The paper is organized as follows: Section \ref{sec:2} devotes for the wave function and surface tension at zero tenperature. The effect from surface excitations to surface tension are presented in Section \label{sec:3}. Conclusions are presented in Section \ref{sec:4}.

\section{Wave function and surface tension at zero temperature\label{sec:2}}

We start by considering a BEC described the Lagrangian density \cite{Pethick2008}
\begin{equation}
{\cal L}=\psi^*(\vec{r},t)\left(-i\hbar\frac{\partial}{\partial t}-\frac{\hbar^2}{2m}\nabla^2\right)\psi(\vec{r},t)-\mu\left|\psi(\vec{r},t)\right|^2+\frac{g}{2}\left|\psi(\vec{r},t)\right|^4,\label{eq:1}
\end{equation}
wherein $\mu$ is the chemical potential, which has the form $\mu=g\rho$ in the mean field theory with $\rho$ the particle density. The field operator $\psi(\vec{r},t)$ depends on both the coordinate $\vec{r}$ and time $t$. The interatomic interaction potential between the atoms can be chosen as the hard-sphere model. In the Born approximation, the strength of the interaction between pairwise atoms is determined via the $s$-wave scattering length $a_s$ as $g=4\pi\hbar^2a_s/m$. Now, thermodynamic stability requires that $g>0$, i.e., the boson interactions are repulsive \cite{Pethick2008,Pitaevskij2010}.

Minimizing this Lagrangian with respect to field operator one arrives at the time-dependent Gross-Pitaevskii equation
\begin{eqnarray}
i\hbar\frac{\partial \psi^*(\vec{r},t)}{\partial t}=-\frac{\hbar^2}{2m}\nabla^2\psi^*(\vec{r},t)+g\left|\psi^*(\vec{r},t)\right|^2\psi^*(\vec{r},t).\label{timeGP}
\end{eqnarray}
Our system exhibits homogeneity in the $(x,y)$-directions but is inhomogeneous in the $z$-direction due to the presence of a hard wall at $z=0$. At zero temperature the condensed part is described as a whole by a wave function $\psi_0(z)$ satisfying
\begin{eqnarray}
\psi_0(z)=\psi(\vec{r},t)e^{\frac{i}{\hbar}\mu t}.\label{wave}
\end{eqnarray}
Plugging (\ref{wave}) into (\ref{timeGP}) yields the time-independent GP equation
\begin{eqnarray}
-\frac{\hbar^2}{2m}\frac{d^2\psi_0(z)}{dz^2}-\mu\psi_0(z)+g\left|\psi_0(z)\right|^2\psi_0(z)=0.\label{notimeGP}
\end{eqnarray}
The boundary condition is imposed that the wave function vanishes at the hard wall
\begin{eqnarray}
\psi_0(0)=0.\label{boundary}
\end{eqnarray}
Using the healing length $\xi=\hbar/\sqrt{2mg\rho}$  one can introduce the dimensionless coordinate $\tilde z=z/\xi$. Therefore the time-independent GP equation (\ref{notimeGP}) and the boundary condition (\ref{boundary}) can be written in dimensionless form
\begin{eqnarray}
-\frac{d^2\tilde\psi_0(\tilde z)}{d\tilde z^2}-\tilde\psi_0(\tilde z)+\tilde\psi_0^3(\tilde z)&=&0,\nonumber\\
\tilde\psi_0(0)&=&0,\label{dimensionless}
\end{eqnarray}
where the normalized wave function is defined as $\tilde\psi_0(z)=\psi_0(z)/\sqrt{\rho}$. The solution for (\ref{dimensionless}) has the well-known form
\begin{eqnarray}
\tilde\psi_0(\tilde z)=\tanh\left(\frac{\tilde z}{\sqrt{2}}\right),\label{solution1}
\end{eqnarray}
or in dimensional form
\begin{eqnarray}
\psi_0(z)=\sqrt{\rho}\tanh\left(\frac{z}{\sqrt{2}\xi}\right).\label{solution0}
\end{eqnarray}

We now calculate the surface tension or wall tension. At zero temperature, the grand potential is read
\begin{eqnarray}
\Omega=\int d^3\vec{r}\left(-\psi_0^*\frac{\hbar^2}{2m}\nabla^2\psi_0-\mu\psi_0^2+\frac{g}{2}\psi_0^4\right),\label{potential}
\end{eqnarray}
or in dimensionless form
\begin{eqnarray}
\Omega={\cal A}P\xi\int d\tilde z\left(\tilde\psi_0^*\frac{d^2\tilde\psi_0}{d\tilde z^2}-\tilde\psi_0^2+\frac{1}{2}\tilde\psi_0^4\right),\label{potential1}
\end{eqnarray}
in which ${\cal A}=\int dxdy$ is the area of the surface and $P=gn^2/2$ is the bulk pressure.

By multiplying both sides of the above Eq. (\ref{dimensionless}) by $d\tilde\psi_0/d\tilde z$ and integrating by parts with respect to $\tilde z$, under the constraint imposed by the boundary condition, one obtains the "constant of motion"
\begin{eqnarray}
\left(\frac{d\tilde\psi_0}{d\tilde z}\right)^2+\tilde\psi_0^2-\frac{1}{2}\tilde\psi_0^4=\frac{1}{2}.\label{motion}
\end{eqnarray}
Combining Eqs. (\ref{motion}) and (\ref{potential1}) one arrives at
\begin{eqnarray}
\gamma(0)=4P\xi\int_0^\infty \left(\frac{d\tilde\psi_0}{d\tilde z}\right)^2d\tilde z.\label{gamma}
\end{eqnarray}
Inserting (\ref{solution1}) into (\ref{gamma}) one obtains the surface tension at zero temperature
\begin{eqnarray}
\gamma(0)=\frac{4\sqrt{2}}{3}P\xi.\label{gamma1}
\end{eqnarray}
Eq. (\ref{gamma1}) shows that the surface tension (or wall tension) is proportional to the healing length. A deviation from this value has been found in some approximations \cite{Indekeu2015,Deng2016}.

It is worthy of note that the quantum fluctuations arising from zero-point energy are neglected in Eq. (\ref{gamma1}), as it has been indicated that their contribution is negligible for a dilute Bose gas \cite{VanThu2022}.

\section{Effect from surface excitations to surface tension\label{sec:3}}

As already mentioned, the surface tension in Eq. (\ref{gamma1}) has been calculated within assumption that all bosonic atoms stay in the ground state at zero temperature. In fact, a number of the atoms is in the excited states. These atoms perform the surface excitation \cite{Pethick2008}.

To analyze the surface excitations, the wave function is expressed as the sum of the condensate wave function, $\psi_0(z)$, and the non-condensed component,
$\delta\psi(\vec{r}_\perp,t)$ \cite{Pethick2008}
\begin{eqnarray}
\psi^*(\vec{r},t)=\left[\psi_0(z)+\delta\psi(\vec{r}_\perp,t)\right]e^{-\frac{i}{\hbar}\mu t},\label{expand1}
\end{eqnarray}
in which $\vec{r}_\perp=(x,y)$ lies in the plane orthogonal to $z$-axis. Substituting (\ref{expand1}) into (\ref{timeGP}) one arrives at the equation for the non-condensed part up to first order
\begin{eqnarray}
i\hbar\frac{\partial\delta\psi(\vec{r}_\perp,t)}{\partial t}=\left(-\frac{\hbar^2}{2m}\nabla^2-\mu+2g\psi_0^2\right)\delta\psi(\vec{r}_\perp,t)+g\psi_0^2\delta\psi^*(\vec{r}_\perp,t).\label{expand2}
\end{eqnarray}
The wave function of the non-condensed part is now disturbed as a sum of plane waves with the wave vector $\vec{k}$ traveling within the surface in $\vec{r}_\perp=(x,y)$-plane \cite{Pitaevskii1961}
\begin{eqnarray}
\delta\psi(\vec{r}_\perp,t)=u_{\vec{k}}(z)\exp(i\vec{k}\vec{r}_\perp-i\omega t)+v^*_{\vec{k}}(z)\exp(-i\vec{k}\vec{r}_\perp+i\omega t),\label{expand2}
\end{eqnarray}
in which $u_{\vec{k}}$ and $v_{\vec{k}}$ are magnitude of the plane wave. The angular frequency $\omega=\omega(k)$ allows us to determine the energy of the surface excitation $\varepsilon_k=\hbar\omega$.
Substituting Eq. (\ref{expand2}) into (\ref{expand1}) and then (\ref{timeGP}) we arrive at the Bogoliubov de Gennes (BdG) equations
\begin{subequations}
\begin{eqnarray}
\left[-\frac{\hbar^2}{2m}\left(\frac{d^2}{dz^2}-k^2\right)-\mu+2g\psi_0^2\right]u_{\vec{k}}+g\psi_0^2v_{\vec{k}}&=&\varepsilon_k u_{\vec{k}},\label{BdG1a}\\
\left[-\frac{\hbar^2}{2m}\left(\frac{d^2}{dz^2}-k^2\right)-\mu+2g\psi_0^2\right]v_{\vec{k}}+g\psi_0^2u_{\vec{k}}&=&-\varepsilon_k v_{\vec{k}}.\label{BdG1b}
\end{eqnarray}\label{BdG1}
\end{subequations}
For the hard wall boundary, the surface excitations have to be satisfied the normalization conditions
\begin{eqnarray}
\int \left(\left|u_{\vec{k}}\right|^2-\left|v_{\vec{k}}\right|^2\right)dz=1.\label{norm}
\end{eqnarray}
Introducing the dimensionless wave vector $\kappa=k\xi$ and two functions
\begin{eqnarray}
\Sigma_{\vec{k}}=u_{\vec{k}}+v_{\vec{k}},\nonumber\\
\Delta_{\vec{k}}=u_{\vec{k}}-v_{\vec{k}},\label{newfunctions}
\end{eqnarray}
the BdG equations (\ref{BdG1}) can be rewritten in matrix form
\begin{eqnarray}
\left(
  \begin{array}{cc}
    0 & {\cal D}_k-{\cal F}_k \\
    {\cal D}_k+{\cal F}_k & 0 \\
  \end{array}
\right)\left(
         \begin{array}{c}
           \Sigma_{\vec{k}} \\
           \Delta_{\vec{k}} \\
         \end{array}
       \right)=\tilde\varepsilon_k\left(
         \begin{array}{c}
           \Sigma_{\vec{k}} \\
           \Delta_{\vec{k}} \\
         \end{array}
       \right),\label{BdG2}
\end{eqnarray}
in which
\begin{eqnarray}
{\cal D}_k&=&-\left(\frac{d^2}{d\tilde z^2}-\kappa\right)-1+2\tanh^2\left(\frac{\tilde z}{\sqrt{2}}\right),\nonumber\\
{\cal F}_k&=&\tanh^2\left(\frac{\tilde z}{\sqrt{2}}\right),\nonumber\\
\tilde\varepsilon_k&=&\frac{\varepsilon_k}{g\rho_0}.\label{DF}
\end{eqnarray}
The boundary condition for $\Sigma_{\vec{k}}$ and $\Delta_{\vec{k}}$ is determined from the boundary at the hard wall. The requirement of the wave function (\ref{solution1}) leads to \cite{Pikhitsa2014}
\begin{eqnarray}
&&\Sigma_{\vec{k}}(0)=\Delta_{\vec{k}}(0)=0,\nonumber\\
&&\Sigma'_{\vec{k}}(0)=0.\label{boundary}
\end{eqnarray}
The BdG equations (\ref{BdG2}) together with boundary condition (\ref{boundary}) are proven yielding two kinds of excitations, which are ripplon or capillary wave and surface phonon \cite{Pikhitsa1992}. The ripplon mode with the dispersion relation
\begin{eqnarray}
\tilde\varepsilon^2_\kappa=\frac{4\sqrt{2}}{3}\kappa^3,\label{ripllon1}
\end{eqnarray}
or in dimensional form
\begin{eqnarray}
\varepsilon_k=\hbar\sqrt{\frac{\gamma(0)}{m\rho_0}}k^{3/2}.\label{ripllon2}
\end{eqnarray}
The second kind is the surface phonon with the dispersion relation
\begin{eqnarray}
\varepsilon_k=\hbar\sqrt{\frac{g\rho_0}{m}}k.\label{phonon}
\end{eqnarray}

We now proceed to examine the contribution of surface excitations to surface tension. As previously stated, these excitations propagate within the surface near the hard wall in the plane $\vec{r}_\perp=(x,y)$. This characteristic implies that the system exhibits a two-dimensional topological singularity. At a given nonzero temperature $T$, the surface tension is expressed as follows \cite{Atkins1953,E.M.Lifshitz1980}
\begin{eqnarray}
\gamma\equiv\gamma(0)+\gamma(T)=\gamma(0)+k_BT\sum_i\int\frac{d^2\vec{k}}{(2\pi)^2}\ln\left(1-e^{-\varepsilon_k^{(i)}/k_BT}\right),\label{gammaT}
\end{eqnarray}
where $k_B$ represents the Boltzmann constant, and $i=\rm r,p$ denotes ripplons and phonons, respectively. The summation accounts for contributions from capillary waves and surface phonons. Due to the two-dimensional topological singularity, the second term on the right-hand side of Eq. (\ref{gammaT}) can be rewritten as
\begin{eqnarray}
\gamma^{(i)}(T)=\frac{k_BT}{2\pi}\int_0^\infty k\ln\left(1-e^{-\varepsilon_k^{(i)}/k_BT}\right).\label{gammaT1}
\end{eqnarray}
For ripplons, substituting Eq. (\ref{ripllon2}) into Eq. (\ref{gammaT1}) yields
\begin{eqnarray}
\gamma^{\rm(r)}(T)=-\frac{\Gamma(7/3)\zeta(7/3)(m\rho)^{2/3}}{4\pi\hbar^{4/3}\gamma(0)^{2/3}}(k_BT)^{7/3}.\label{ripplon3}
\end{eqnarray}
Eq. (\ref{ripplon3}) indicates that capillary waves cause the surface tension to decrease with increasing temperature, following a power-law dependence. Similarly, for surface phonons, substituting Eq. (\ref{phonon}) into Eq. (\ref{gammaT1}) results in
\begin{eqnarray}
\gamma^{\rm(p)}(T)=-\frac{\zeta(3)m}{2\pi\hbar^2g\rho}(k_BT)^3.\label{phononnew}
\end{eqnarray}
\begin{figure}[t]%% placement specifier
\centering%% For centre alignment of image.
\includegraphics[width = 0.6\linewidth]{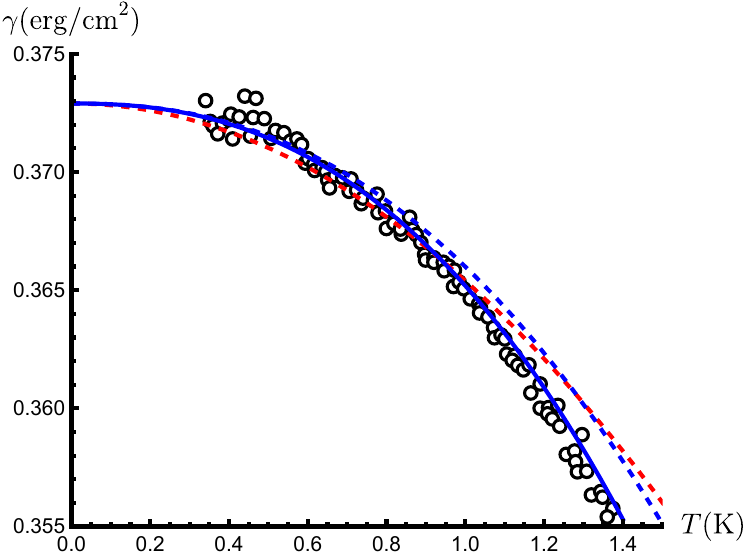}
% Use \caption command for figure caption and label.
\caption{(Color online) The evolution of the surface tension of helium 2 as a function of temperature. The dashed red and dashed blue lines correspond to results of Singh \cite{Singh1962} and Atkins \cite{Atkins1953}. Our result is shown by the solid blue line and the experimental data \cite{Atkins1965} are marked by the open circle dots.}\label{fig1}
\end{figure}
It is evident that the surface tension decreases further when the contribution of surface phonons is taken into account. In summary, the surface tension at finite temperature can be expressed from Eqs. (\ref{gamma1}), (\ref{ripplon3}) and (\ref{phononnew}) as follows
\begin{eqnarray}
\gamma=\frac{4\sqrt{2}}{3}P\xi-\frac{\Gamma(7/3)\zeta(7/3)(m\rho)^{2/3}}{4\pi\hbar^{4/3}\gamma(0)^{2/3}}(k_BT)^{7/3}-\frac{\zeta(3)m}{2\pi\hbar^2g\rho}(k_BT)^3.\label{total}
\end{eqnarray}

In the study of dilute BEC, an important parameter frequently employed in calculations is the gas parameter, defined as $\alpha_s=\rho a_s^3$. The condition for a system to be considered dilute is given by $\alpha_s\ll1$ \cite{Andersen2004}. Neglecting the temperature dependence of the chemical potential, Eq. (\ref{total}) can be rewritten in a conventional form as
\begin{eqnarray}
\frac{\gamma-\gamma(0)}{\gamma(0)}=-\left(\frac{243}{4}\right)^{1/3}\sqrt{\pi}\Gamma(7/3)\zeta(7/3)\alpha_s^{1/2}\left(\frac{k_BT}{\mu}\right)^{7/3}-6\sqrt{\pi}\zeta(3)\alpha_s^{1/2}\left(\frac{k_BT}{\mu}\right)^3,\label{totalns}
\end{eqnarray}
or, equivalently, in numerical form
\begin{eqnarray}
\frac{\gamma-\gamma(0)}{\gamma(0)}=-11.74\alpha_s^{1/2}\left(\frac{k_BT}{\mu}\right)^{7/3}-12.78\alpha_s^{1/2}\left(\frac{k_BT}{\mu}\right)^3.\label{totalns1}
\end{eqnarray}
Due to the extremely small magnitude of the gas parameter, the contribution of temperature to the surface tension is negligible. Moreover, in the most BEC experiments, $k_BT$ is significantly smaller than the chemical potential \cite{Pitaevskij2010}, further diminishing this contribution. For instance, in a recent experiment \cite{Mordini2020}, a BEC was generated using the sodium-23 isotope, with $\alpha_s\approx 2\times 10^{-6}$, while the temperature was controlled within the range of 150 nK to 280 nK. Under these conditions, Eq. (\ref{totalns1}) yields
\begin{eqnarray}
\frac{\left|\gamma-\gamma(0)\right|}{\gamma(0)}\leq 0.7\%.\label{totalns2}
\end{eqnarray}
This result provides unequivocal evidence that the influence of nonzero temperature on the surface tension of a BEC is very small.

We now compare our theoretical results with experimental data on surface tension. Due to its significant applications, the surface tension of liquid helium II has attracted considerable attention from experimentalists. The first experimental study was conducted by Allen and Misener \cite{Allen1938}, who measured the surface tension of helium II within a high-temperature range (from 1 K to 5 K). To explain the temperature dependence of surface tension, Atkins \cite{Atkins1953} proposed that the normal modes of a liquid helium surface could be interpreted as surface tension waves, leading to a temperature-dependent decrease in surface tension following a power-law behavior
\begin{eqnarray}
\gamma=\gamma(0)-6.9\times 10^{-3}T^{7/3},\label{Atkins}
\end{eqnarray}
where $\left[\gamma\right]={\rm erg/cm}^{-2}$ and $\left[T\right]={\rm K}$. In 1962, Singh \cite{Singh1962} proposed a simplified model based on the ideal gas approximation for liquid helium II, deriving the following relation
\begin{eqnarray}
\gamma=\gamma(0)-\frac{\pi m\zeta(2)}{8\pi^2\hbar^2}(k_BT)^2.\label{Singh}
\end{eqnarray}
Using the conventional capillary-rise method, Atkins and Narahara \cite{Atkins1965} measured the surface tension of liquid helium II in the sufficiently low-temperature region (below 1.4 K) and obtained an empirical relation
\begin{eqnarray}
\gamma=0.372-0.0081 T^{2.5\pm 0.2}~(\rm erg/cm^{2}).\label{Narahara}
\end{eqnarray}

Applying the above experimental results from Ref. \cite{Atkins1965} to our theoretical expression in Eq. (\ref{total}), the surface tension as a function of temperature can be expressed as
\begin{eqnarray}
\gamma=0.3729-0.0069T^{7/3}-7.898\times 10^{-4}T^3~(\rm erg/cm^{2}),\label{result}
\end{eqnarray}
where the surface tension is measured in units of erg/cm$^2$ and the temperature in kelvin. The variation of the surface tension of helium II as a function of temperature is illustrated in Fig. \ref{fig1}. The experimental data obtained by Atkins and Narahara \cite{Atkins1965} are represented by open circular markers. The dashed red line corresponds to the results of Singh {\it et al.} \cite{Singh1962}, while the dashed blue curve represents the predictions of Atkins \cite{Atkins1953}. Our theoretical result in Eq. (\ref{result}) is depicted by the solid blue line, demonstrating excellent agreement with the experimental data. Indeed, fitting our result to the empirical form in Eq. (\ref{Narahara}) yields
\begin{eqnarray}
\gamma=0.372897-0.00773143T^{2.44003}~ (\rm erg/cm^{2}).\label{resultfit}
\end{eqnarray}
This result is slightly different from (\ref{Narahara}). 
It is worth noting that, in addition to ripplons and phonons, surface excitations in liquid helium II also include contribution from rotons \cite{Brouwer1967}, characterized by the dispersion relation
\begin{eqnarray}
\varepsilon_k^{\rm(ro)}=\Delta +\frac{\hbar^2(k^2-k_0^2)}{2m^*},\label{roton}
\end{eqnarray}
where $\Delta$ represents an energy gap and $m^*\approx (0.16\pm0.01)m$. Substituting Eq. (\ref{roton}) into Eq. (\ref{gammaT1}) gives \cite{Brouwer1967}
\begin{eqnarray}
\gamma^{\rm (ro)}=\pm 0.031 e^{-\Delta/k_BT}T^{3/2}~ (\rm erg/cm^{2}).
\end{eqnarray}
For liquid helium II, the energy shift is approximately $\Delta/k_B\approx 8.65\pm0.04$ K, $k_0=1.02\pm0.01$ \AA$^{-1}$ \cite{Huang2001}. Naturally, the contribution of rotons is extremely small due to the presence of the exponential term.

\section{Conclusion\label{sec:4}}

In the preceding sections, the influence of surface excitations on the surface tension of a single BEC confined by a hard wall has been investigated. These surface excitations comprise ripplons and phonons propagating along the bounded surface. When applied to liquid helium II, an empirical relation for surface tension as a function of temperature is derived, exhibiting close agreement with experimental data. A comparison with previous results has been conducted, demonstrating the superiority of our empirical relation.

For a single BEC, our analysis indicates that the contribution of nonzero temperature to surface tension is less than 0.7\% in recent experiments. Notably, in our calculations, the contribution of the zero-point energy of surface excitations has been neglected. Taking into account the zero-point energy, which will cause the quantum fluctuations. In that case, the pressure in the surface tension at zero temperature becomes \cite{VanThu2022}
\begin{eqnarray}
P=\frac{1}{2}g\rho^2\left(1+\frac{128}{15\sqrt{\pi}}\alpha_s^{1/2}\right).\label{pressure}
\end{eqnarray}
Nonetheless, this contribution is sufficiently small.

 Regarding the wetting phenomenon in BECs, the interface tension between components and/or wall tension must be considered \cite{VanSchaeybroeck2008}.

\begin{acknowledgements}
This research is funded by Vietnam National Foundation for Science and Technology Development (NAFOSTED) under grant number 103.01-2023.12.
\end{acknowledgements}

\section*{Conflict of interest}
All of the authors declare that we have no conflict of interest.

\bibliography{tension.bib}% Produces the bibliography via BibTeX.

\end{document}